\documentclass[final,3p,11p]{elsarticle}

\usepackage{graphicx}
\usepackage[all]{xy}

\usepackage{amssymb}

\usepackage{hyperref}
\hypersetup{
    bookmarks=true,         
    unicode=false,          
    pdftoolbar=true,        
    pdfmenubar=true,        
    pdffitwindow=false,     
    pdfstartview={FitH},    
    pdftitle={My title},    
    pdfauthor={Author},     
    pdfsubject={Subject},   
    pdfcreator={Creator},   
    pdfproducer={Producer}, 
    pdfkeywords={keyword1} {key2} {key3}, 
    pdfnewwindow=true,      
    colorlinks=false,       
    linkcolor=red,          
    citecolor=green,        
    filecolor=magenta,      
    urlcolor=cyan           
}

\def\eqref#1{Eq.~(\ref{eq:#1})}
\def\eqlab#1{\label{eq:#1}}
\def\figref#1{Fig.~(\ref{fig:#1})}
\def\figlab#1{\label{fig:#1}}
\def\tabref#1{Tab.~(\ref{tab:#1})}
\def\tablab#1{\label{tab:#1}}

\newcommand{\be}{\begin{equation}}
\newcommand{\ee}{\end{equation}}
\newcommand{\bea}{\begin{eqnarray}}
\newcommand{\eea}{\end{eqnarray}}

\journal{Physics Letters B}

\begin{document}

\begin{frontmatter}

\title{The impact of energy conservation in transport models on the $\pi^-/\pi^+$ multiplicity ratio in heavy-ion collisions and the symmetry energy}
\author{M.D. Cozma}
\ead{dan.cozma@theory.nipne.ro}
\address{IFIN-HH, Reactorului 30, 077125 M\v{a}gurele-Bucharest, Romania}

\begin{abstract} 
The charged pion multiplicity ratio in intermediate energy central heavy-ion collisions has been proposed 
as a suitable observable to constrain the high density dependence of the isovector part of the equation of state. A
comparison of various transport model predictions with existing experimental data has led, however, 
to contradictory results. Using an upgraded version of the 
T\"ubingen QMD transport model, which allows the conservation of energy at a local or global level by accounting
for the potential energy of hadrons in two-body collisions and leading thus to particle production threshold shifts, 
we demonstrate that compatible constraints for the symmetry energy stiffness can be extracted from pion multiplicity
and elliptic flow observables. However, pion multiplicities and ratios are proven to be highly sensitive to the yet unknown
isovector part of the in-medium $\Delta$(1232) potential which hinders, at present, the extraction of meaningful information on
the high density dependence of the symmetry energy. A solution to this problem together with the inclusion of contributions presently neglected, 
such as in-medium pion potentials and retardation effects, are needed for a final verdict on this topic. 

\end{abstract}


\end{frontmatter}

\section{Introduction}
\label{introduction}
The isovector part of the equation of state of nuclear matter (asy-EoS), commonly known as symmetry energy (SE),
has an important impact on the structure of rare isotopes, dynamics and spectra of heavy-ion collisions 
and on certain astrophysical  processes such as neutron star cooling, their chemical composition, 
and supernovae explosions~\cite{Baran:2004ih,Li:2008gp}. While its dependence on density close to
the saturation point ($\rho_0$) has been reliably constrained by nuclear structure and intermediate energy
nuclear reaction experiments~\cite{Horowitz:2014bja,Vinas:2013hua},
its behavior at densities close to or above 2$\rho_0$ is currently still uncertain. 
It is thus of uttermost importance to study experimentally nuclear matter at suprasaturation densities, a regime reached
in Earth based laboratories only during the process of heavy-ion collisions (HICs). To this end several promising observables have been
identified: the ratio of neutron/proton yields of squeezed out nucleons~\cite{Yong:2007tx}, 
light cluster emission~\cite{Chen:2003qj}, $\pi^-/\pi^{+}$ multiplicity ratio (PMR) in central collisions~
\cite{Xiao:2009zza,Feng:2009am,Xie:2013np}, elliptic flow related observables~\cite{Li:2002qx} and others. 

Constraints for the asy-EoS stiffness from the elliptic flow ratio of neutrons vs. hydrogen and neutrons vs. protons have been
extracted using the UrQMD transport model and power-law~\cite{Russotto:2011hq} and contact Skyrme
interactions~\cite{Wang:2014rva} parametrizations of the symmetry potential with the results for the slope parameter of SE at saturation
 $L$=83$\pm$52 MeV and $L$=89$\pm$45 MeV respectively. A similar analysis making use of the T\"ubingen
QMD (T\"uQMD) transport model and the Gogny parametrization of the symmetry potential~\cite{Cozma:2013sja}
and using the experimental data for either elliptic flow difference or ratio of neutrons and protons
has resulted in a higher average value for the slope parameter of the asy-EoS: $L$=123$^{+33}_{-52}$ MeV. 
Very similar constraints for the stiffness can be extracted using the 
IBUU transport model too~\cite{Yong:2013rva}. In all cases reanalyzed sets of the 90's experimental data for $^{197}$Au+$^{197}$Au
collisions at an incident projectile energy of 400 MeV/nucleon, obtained by the
FOPI-LAND collaboration~\cite{Leifels:1993ir,Lambrecht:1994cp}, have been employed.

Attempts to constrain the stiffness of asy-EoS making use of the FOPI experimental~\cite{Reisdorf:2006ie} data for the $\pi^-/\pi^{+}$ 
multiplicity ratio (PMR) have resulted in a series of contradictory results: Xiao $et\,al.$~\cite{Xiao:2009zza} 
made use of the IBUU transport model supplemented by the isovector momentum dependent Gogny inspired 
parametrization of SE~\cite{Das:2002fr} to point toward a soft asy-EoS, the study of Feng and Jin~\cite{Feng:2009am}, 
which employed IQMD and a power-law parametrization of the potential part of the SE, favors a stiff asy-EoS and lastly
 Xie $et\,al.$~\cite{Xie:2013np} addressed the same issue within the Boltzmann-Langevin approach and a 
power-law parametrization of asy-EoS presenting support for a super-soft scenario for the SE. 
 Still, the recent study of Hong $et\,al.$~\cite{Hong:2013yva} presents the claim that the  
$\pi^-/\pi^{+}$ multiplicity ratio is insensitive to the slope of SE at saturation, which is suggested to may be due to
the inclusion of the S-wave pion potential, a feature that models previously employed in computing the PMR do not share. 
Additionally, the extracted stiffness from elliptic flow and pion multiplicity ratios data are often in contradiction to each other 
for a given transport model.

The described state of affairs on the theoretical side is clearly unacceptable. In view of the upcoming new experimental measurements
(ASYEOS collaboration~\cite{Russotto:2014zba} for elliptic flow and SAMURAI TPC collaboration~\cite{Shane:2014tsa} for PMR), the observed 
large discrepancy between the asy-EoS stiffness extracted from elliptic flow versus charged pion multiplicity
ratio data has to be eliminated. Previous attempts, which focused on the impact of in-medium modification of the
pion-nucleon interaction~\cite{Xu:2013aza,Xu:2009fj} or the kinetic part of SE~\cite{Li:2014vua}, 
size of the neutron-skin thickness~\cite{Wei:2013sfa} and threshold modifications due to the inclusion of self-energy 
effects~\cite{Ferrini:2005jw,Ferini:2006je,Song:2014xza} on the PMR value have usually reported a small effect and not always in the
right direction.

Transport models that employ momentum and/or isospin dependent mean fields do not conserve the total energy at an
event by event basis. Even the accepted wisdom in the field, that energy is conserved ``on average'', is shown to be inaccurate
by a value much larger than the pion rest mass. Consequently, an upgrade of the T\"ubingen QMD transport model, that alleviates
this problem, has been developed and its relevant details are presented in this Letter (Section 2), together with the impact on 
pion multiplicities and ratios and briefly on the neutron-proton elliptic flow ratio (Section 3).
Additionally, the impact of poorly known quantities, 
in particular the isovector part of the in-medium $\Delta$(1232) potential, on the value of PMR is studied
and conclusions upon the feasibility of using it, given the current knowledge, for constraining the high density 
dependence of the symmetry energy are drawn.

\section{The model}
\label{themodel}
Heavy-ion collisions have been simulated by using the QMD transport model developed
in T\"{u}bingen~\cite{Khoa:1992zz,UmaMaheswari:1997ig}. The model has been previously applied to the study of
dilepton emission in HICs~\cite{Shekhter:2003xd,Cozma:2006vp,Santini:2008pk}, stiffness 
of the equation of state of symmetric nuclear matter~\cite{Fuchs:2000kp} and various in-medium effects relevant 
for the dynamics of HICs~\cite{UmaMaheswari:1997ig,Fuchs:1997we}. It has been previously upgraded to 
accommodate density dependent cross-sections and an isospin dependent EoS~\cite{Cozma:2011nr}.
For the present study the Gogny inspired momentum dependent parametrization of the isovector
part of the equation of state~\cite{Das:2002fr} has been selected. It contains a  parameter denoted $x$ which has been
introduced to allow adjustments of the stiffness of the asy-EoS, negative and positive values corresponding to a stiff
and a soft density dependence of the symmetry energy, respectively. 
Corresponding values for the slope parameter $L$ and curvature $K_{sym}$
coefficients for the values of $x$ encountered in this study are listed in~\tabref{xvslsymksym}. 
With this choice for the parametrization of SE the parameter space to be constrained is given by the pair ($S_0$, $x$).
It is equivalent to the more familiar one used in a similar context to analyze nuclear structure
experimental data, ($S_0$, $L$), since the impact of the asy-EoS on observables originates predominantly
from density regions close to saturation. When employing transport models to study
HICs it is however customary to additionally fix the magnitude of SE at saturation
($S_0$)~\cite{Xiao:2009zza,Feng:2009am,Xie:2013np,Russotto:2011hq,Wang:2014rva,Hong:2013yva} reducing the probed parameter space
to a one dimensional one, an approach adopted also in this study by setting $S_0$=30.6 MeV (and thus independent of $x$)~\cite{Das:2002fr}. 

\begin{table}
\centering
\begin{tabular}{|c|c|c|}
\hline\hline
x&L(MeV)&K$_{sym}$(MeV)\\ 
\hline\hline
-2& 152 & 418 \\
-1& 106 & 127 \\
0& 61 & -163\\
1& 15 & -454\\
2& -31& -745\\
\hline\hline
\end{tabular}
\caption{Values for $L$ and $K_{sym}$ coefficients appearing in the Taylor expansion of the symmetry energy around saturation density,
$S(\rho)=S_0+L/3\,u+K_{sym}/18\,u^2+\dots$ with $u=\frac{\rho-\rho_0}{\rho_0}$ and $S_0$=30.6 MeV, for given values of the stiffness parameter $x$.}
\tablab{xvslsymksym}
\end{table}

The strength of baryon resonances  potentials in nuclear matter is presently uncertain at best.
Results for the isoscalar part of the $\Delta$(1232)-potential differ considerably, depending on the source.
Phenomenological studies of inclusive electron-nucleus (He, C, Fe) scattering data favor an attractive potential,
deeper than that of the nucleon-nucleus system~\cite{O'Connell:1990zg}. 
Ab-initio calculations, using as input well established microscopical nucleon-nucleon potentials (Argonne $v_{28}$)
within the framework of the Bethe-Brueckner-Goldstone method~\cite{Baldo:1994fk} 
or one-boson exchange nucleon-nucleon potentials in the relativistic Dirac-Brueckner model, which allow also a good reproduction
of the elastic pion-nucleon $P_{33}$ phase-shift~\cite{deJong:1992wm}, do however arrive at a mildly repulsive
isoscalar $\Delta$-potential. The latter result is due to dominant repulsive contributions of total isospin I=2, a channel which cannot be
sufficiently constrained by elastic nucleon-nucleon scattering data. In view of these results it is customary to choose,
 in transport models, the isoscalar component of baryon resonances potential equal to that of the nucleon, a choice adopted here too. 

To our best knowledge, there have been no attempts reported in the literature to extract information on the isovector part of the
 $\Delta$(1232) potential. The choice most often employed for isospin 3/2 resonances is guided by the decay 
branching ratios of the isospin quadruplet components into the possible pion-nucleon pairs~\cite{Li:2002yda}, leading to
\begin{eqnarray}
 \begin{array}{lcrcr}
V_{\Delta^-}&=& V_n&&\\
V_{\Delta^0}&=&(2/3)\,V_n&+&(1/3)\,V_p\\
V_{\Delta^+}&=&(1/3)\,V_n&+&(2/3)\,V_p\\
V_{\Delta^{++}}&=&&&V_p
\eqlab{choicedeltapot}
\end{array}
\end{eqnarray}
while for the isospin 1/2 resonances the potentials of the isospin partners are taken the same as those of the corresponding component
of the nucleon isospin doublet.

The enforcement of the principle $action=reaction$ allows energy conservation at an event by event basis if the transport
model is employed in the Vlasov mode. For that to be true when the collision term is switched on one has to take into account
the nuclear matter potential energy of each particle in the process of determining the kinematical variables of the particles
 in the final state of a collision. To be precise, the in-vacuum energy conservation relation
\begin{eqnarray}
 \sum_i \sqrt{p_i^2+m_i^2}&=&\sum_j \sqrt{p_j^2+m_j^2},
\end{eqnarray}
where indexes i and j run on the particles involved in the collision (or decay, reabsorption), has to be replaced by
\begin{eqnarray}
 \sum_i \sqrt{p_i^2+m_i^2}+U_i&=&\sum_j \sqrt{p_j^2+m_j^2}+U_j,
\end{eqnarray}
where now both indexes run over all the particles of the whole system present at the moment of the collision. This
relation enforces energy conservation at a global level (GEC). The restriction of the relation above only to the particles
involved in the collision will be called in the following local energy conservation (LEC), as only the total energy of the colliding
particles is conserved in this case (global energy conservation is fulfilled also in this case if the mean-field potentials are momentum
and isospin independent). In this case and with the simplifying assumption of momentum independent mean-field potentials 
the multiplicities of produced $\Delta^{-},\Delta^{0}(1232)$ isobars increase with respect to the free case, 
the opposite holding true for the $\Delta^{+},\Delta^{++}(1232)$ iso-quadruplet partners. This effect is a direct consequence 
of the attractive/repulsive nature of the proton/neutron isovector
potentials. Consequently, an enhancement of the PMR is expected, a conclusion that can however be altered in the realistic case
of a momentum dependent symmetry potential. The situation when potential energy effects are neglected, as was the case with previous versions
of the model, will be referred to in the following as in-vacuum energy conservation (VEC).

The time dependence of the average (over events) potential, kinetic and total energies for the three energy conservation scenarios, during heavy-ion
collisions, is presented in~\figref{enbalance}. As advertised, the GEC scenario (solid curve) allows the conservation of the total energy exactly. This happens
only partially so for the VEC (dash-dotted curve) and LEC (dashed curve) scenarios, LEC being a clear improvement over VEC, about halfway between the VEC and GEC results.
Investigation of the potential and kinetic energy components separately leads to the conclusion that most of the energy conservation violation
for the VEC and LEC scenarios is to be found as a kinetic energy excess in the final particle spectra, and arises from the collision term during
the high-density phase of the reaction. It amounts to about 3 GeV for the case presented here, which corresponds, on average, to about 7.5 MeV
of extra kinetic energy per nucleon. While this value is probably too small to have an observable impact on free nucleon spectra it may have an
impact on the light fragments ones when a theoretical description is attempted using coalescence algorithms that take into account 
the binding energy of clusters. The situation is even more worrying-some for the case of pion production given the large ratio between
the observed energy excess and the pion rest mass and also the accepted mechanism for pion emission close to or below threshold.

\begin{figure}[t]
\centering\includegraphics[width=14pc]{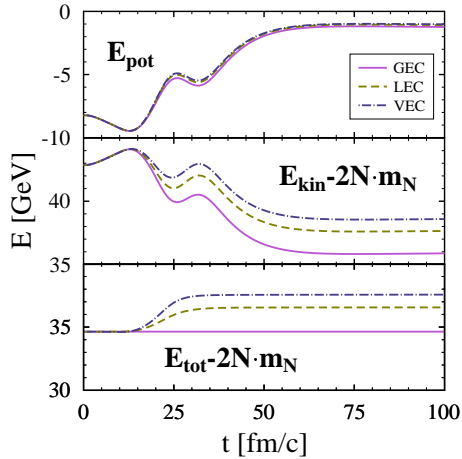}
\caption{\figlab{enbalance} Time dependence of the average potential, kinetic and total energies (from top to bottom)
in central Au+Au collisions at an impact energy of 400 MeV/nucleon for the three energy conservation scenarios discussed in the text.
Contributions due to all particles present at a given moment are included (nucleons, resonances, pions), the rest mass of the nucleons 
is however subtracted, where indicated, for convenience. The asy-EoS
stiffness parameter has been set to x=-2, however the dependence on this parameter is small, being visible, at the 10\% level, only in the relative
difference between LEC and GEC  versus VEC and GEC scenarios. The variance of the presented average values varies between 0.1 and 0.3 GeV, 
the larger value occurring during the maximum compression stage of collisions.}
\end{figure}

Within the GEC (and also LEC) scenario two-body collisions lose, due to potential energy exchange with the fireball, their local character and 
can be viewed as part of a multi-hadron collision. 
In a Feynman diagrammatic like picture the interaction between the baryon pair and the rest of the fireball consists from a 
summation of initial- and final-state interaction contributions supplemented by more complicated terms (so called rescattering terms). 
Only the initial- and final-state interaction contributions will be accounted for in this study.
In a microscopical model these two terms are added coherently leading, besides the individual contributions, to an interference term.
Lacking the knowledge of the quantum scattering amplitudes, we face the impossibility of computing such interference terms and
are hence forced to make some approximations. For elastic nucleon-nucleon scattering 
the difference between the final and initial invariant masses, $\sqrt{s_f}$ and $\sqrt{s_i}$, is on average a few MeV in 
absolute value (bottom panel of~\figref{delsrt}). Consequently the initial and final-state
interaction terms are comparable in magnitude. The elastic nucleon-nucleon collisions are thus evaluated at an energy 
$\sqrt{s^*}$=0.5$(\sqrt{s_i}+\sqrt{s_f})$ assuming a linear dependence of cross-sections on energy in this narrow interval.

For processes that lead to resonance excitation/absorption the situation is different. Owing to the magnitude of energy shifts 
between initial and final states (top and middle panels of~\figref{delsrt}) and the rapid variation with energy of inelastic cross-sections close to threshold 
the initial- (final-) state terms are dominating for the resonance excitation (absorption) process. Consequently the smaller 
contribution is neglected, an approximation that has been estimated, by inspecting the energy dependence of inelastic cross-sections,
to be reasonable up to an incident energy of the projectile 
nucleus $T_{lab}$=800 MeV/nucleon. The energy values at which cross-sections are evaluated are thus chosen to be
$\sqrt{s^*}=\sqrt{s_f}$ and $\sqrt{s^*}=\sqrt{s_i}$ for the resonance excitation and absorption processes, respectively. 
In particular, in the detailed balance formula~\cite{Danielewicz:1991dh},
\begin{eqnarray}
 &&\frac{d\sigma^{(NR\rightarrow NN)}}{d\Omega}=\frac{1}{4}\frac{m_R\,p_{NN}^2}{p_{NR}}\frac{d\sigma^{(NN\rightarrow NR)}}{d\Omega}\times \\
  &&\qquad\quad\bigg(\frac{1}{2\pi}\int_{m_N+m_\pi}^{\sqrt{s_i}-m_N}dM M\,p'_{NR}\,A_R(M)\bigg)^{-1}, \nonumber
\end{eqnarray}
the momenta $p_{NN}$ and $p_{NR}$ have to be evaluated using the invariant masses of the $NN$ (final) and $NR$(initial)
states respectively. Such a prescription can be understood since, in the expression for the cross-section of a 2-body reaction
$NR\rightarrow NN$, $p_{NR}$ originates from the evaluation of the incoming flux, while $p_{NN}$ arises from the final-state phase space.
For the resonance excitation reaction $NN\rightarrow NR$ the situation is obviously reversed. Decay and absorption processes are
treated in a similar manner.

\begin{figure}[t]
\centering\includegraphics[width=15pc]{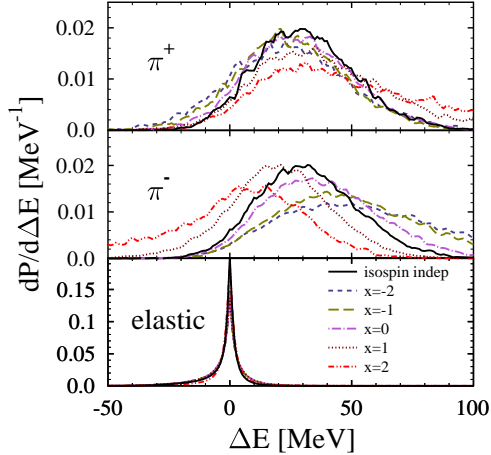}
\caption{\figlab{delsrt} Distribution of the threshold shifts in ${\it succesful}$ two-body elastic $NN\rightarrow NN$ (bottom panel)
and inelastic $NN\rightarrow NR$ (top and middle panels)
nucleon-nucleon collisions for the GEC scenario for the case of an isospin independent interaction (solid curves) and for the isospin dependent
one for various values of the stiffness parameter $x$. For the inelastic channels
labeled as $\pi^+$ (top) and $\pi^-$ (middle) the various contributing reactions are added with weights equal to the branching ratios of each resonance
into the specified pion charge state. The elastic channels are summed up together. The quantity $\Delta$E=$\sqrt{s_f}$-$\sqrt{s_i}$
represents the shift in energy between the invariant masses of the initial and final states. 
The maxima for the elastic scattering reactions distributions occur in the interval -3.0 MeV$\leq\Delta$E$\leq$-1.0 MeV.
Positive values for $\Delta$E correspond to an effective lowering of the threshold position with respect to its vacuum one. }
\end{figure}


In the actual calculations $\sqrt{s_f}$ has to be determined by imposing energy conservation, in the center of mass
frame of the colliding nuclei, for all collision, decay and absorption processes. 
The angular dependence of cross-sections is implicitly (slightly) modified in the iteration process of determining 
the correction to the total energy which originates from the change of potential energy between initial and final 
states and which leads to a different boost between the center of mass frame of the nuclei and the ``modified'' 
center of mass frame of the colliding baryons (in which the vacuum angular
dependence of cross-section is used). In the case of momentum dependent potentials the determination of the allowed ranges for resonance's
masses in the final state of $NN\rightarrow NR$ reactions becomes more laborious. The maximum allowed value of resonance's 
mass (M$_{max}^{LEC,GEC}$) does not appear at zero relative momentum as is the case for the VEC scenario. 
However, for the case of the Gogny potential, $|$M$_{max}^{LEC,GEC}$-M(k=0)$|\leq$0.5 MeV and consequently
the approximation M$_{max}^{LEC,GEC}$=M(k=0) is a very good one due to the presence of Fermi motion, allowing
a considerable speed up of simulations; M(k=0) is the resonance's mass for the case of a zero relative momentum in the state (final or initial) 
which contains a resonance.


\section{Results and Discussion}

One of the aims of the present study is to prove that compatible constraints for the stiffness of asy-EoS can be extracted from
elliptic flow and charged pion multiplicity ratio experimental data together with a good description of individual elliptic flow and
pion multiplicity values. To achieve this last goal (and only for it) certain model parameters will be varied within reasonable limits 
and in-medium modified values for both elastic and inelastic nucleon-nucleon scattering cross-sections will have to be adopted. 
This analysis has been performed for the GEC scenario since, within the current model, it represents the closest approximation
to the real situation: it does conserve total energy but it violates causality since retardation effects and, additionally, the impact
of the pion potential has been neglected. 

For the standard soft value of the compressibility modulus K=210 MeV experimental values for $\pi^-$ and $\pi^+$ multiplicities~\cite{Reisdorf:2006ie}
in $^{197}$Au+$^{197}$Au collisions at incident energies of 400 MeV/nucleon are overestimated by about 20$\%$ and 40$\%$ respectively. Choosing a stiffer
EoS results in fewer nucleon-nucleon collisions and thus pion multiplicities decrease. At values of the compressibility modulus close to K=310 MeV
the best description of the FOPI data is achieved, however the $\pi^-$ and $\pi^+$ multiplicities cannot be equally good described for the same
value of K. Additionally, for such a stiff EoS elliptic flow strength of protons is overestimated independently of the asy-EoS stiffness.
This suggests the need for an in-medium modification of the inelastic channel cross-sections. It was found that the choice K=245 MeV coupled with
medium modifications of both elastic and inelastic channels cross-sections allows a good reproduction of experimental pion multiplicities
together with a fair one  (within 15-20$\%$)for elliptic flow values. For the latter ones an improvement of the description can be achieved by choosing an optical
potential that becomes more repulsive at higher momenta~\cite{Hartnack:1994zz}.

In this work, the medium modification of elastic and inelastic cross-sections was introduced via the effective mass scaling scenario of 
Ref.~\cite{Schulze:1997zz,Persram:2001dg,Li:2005jy} which assumes that in-medium elastic cross-sections scale with the in-medium
masses of the considered particles, an assumption which was extended here also to the inelastic channels
\begin{eqnarray}
 \sigma^{*\,(12\rightarrow 34)}&=&\Big[\frac{m_1^*}{m_1} \frac{m_2^*}{m_2} 
\frac{m_3^*}{m_3} \frac{m_4^*}{m_4}\Big]^{1/2}\,\sigma^{(12\rightarrow 34)},
\eqlab{medcsec}
\end{eqnarray}
the starred quantities representing in-medium values. The density dependence of masses is determined from the expression 
$m^*/m=1/(1+m/p\cdot dU/dp)$. Within this scenario there are no free parameters that are adjusted, 
the density dependence of baryon masses (and hence cross-sections) is fully fixed with the choice of the mean-field potential.

The impact of medium modified cross-sections on pion multiplicities (PM) and pion multiplicity ratio (PMR)
in central (b=0.0 fm) $^{197}$Au+$^{197}$Au collisions at a projectile energy of 400 MeV/nucleon are presented 
in~\figref{pionmult_divecs} and~\figref{pionrat_divecs} respectively. PM are overestimated if vacuum cross-sections are used (short-dashed curves). 
The inclusion of medium effects on cross-sections through the Ansatz in~\eqref{medcsec} on both the elastic and inelastic channels
results in lower PM values (solid curves) that are in good agreement with the FOPI experimental data, most of the effect originating from the modifications
imposed on to the inelastic channels. The impact of medium effects on PMR is however small (compare short-dashed and solid curves
in \figref{pionrat_divecs}) at this impact energy in contrast with results at much lower energy where the impact of the
symmetry energy and medium modifications of cross-sections on
PMR values may be of comparable magnitude~\cite{Guo:2014usa}. Choosing a different optical potential (as for example the ones
used in Ref.~\cite{Hartnack:1994zz}) generally results in variations of PM at the 10-20$\%$ level for both the vacuum and in-medium 
cross-sections scenarios, however the impact on PMR is again small. The impact of the Coulomb interaction is important, neglecting 
it results in an increase of PM by about 20$\%$ and 40$\%$ for the $\pi^-$ and $\pi^+$ mesons respectively, leading to a decrease of the PMR by 
20$\%$ for the GEC scenario.

\begin{figure}[t]
\centering\includegraphics[width=14pc]{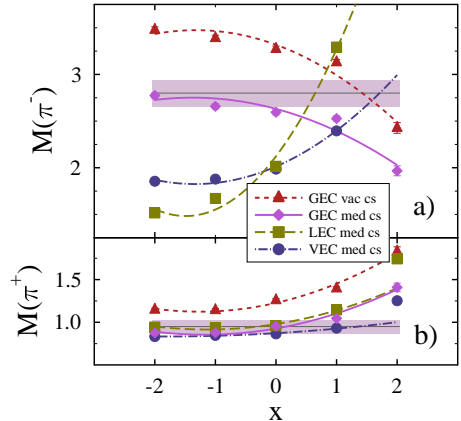}
\caption{\figlab{pionmult_divecs} Multiplicities of $\pi^-$ (top panel) and $\pi^+$ (bottom panel) mesons for the vacuum (``vac cs'') and
in-medium (``med cs'') cross-sections scenarios for the case of global energy conservation (GEC) as a function of the asy-EoS stiffness parameter $x$. 
Results for the LEC and VEC approximations making use of in-medium modified cross-sections (``med cs'') are also displayed.
The experimental FOPI results~\cite{Reisdorf:2006ie} are represented by horizontal bands.}
\end{figure}

A comparison of PM for the VEC, LEC and GEC scenarios is presented in~\figref{pionmult_divecs}. 
The results for the LEC (long-dashed curves) and VEC (dash-dotted curves) scenarios have been obtained by applying 
the corresponding energy conservation constraints, 
without any change of model parameters as compared to the GEC (solid curves) case. In all three cases the medium modified cross-sections
have been used above pion production threshold, while below this point only the inelastic channels have been modified.
It is seen that the $\pi^+$ multiplicities change only for very soft choices of the asy-EoS. The difference between the three energy conservation
scenarios is however dramatic for the $\pi^-$. The multiplicity dependence on the stiffness of the asy-EoS, for the GEC scenario,
can be understood from the magnitude of the threshold shifts for each particular case as presented in~\figref{delsrt}, in particular
the dependence of the location of the peak of the threshold shifts distribution on the stiffness of the asy-EoS, 
for each charge state of the pion separately (top and middle panels for the $\pi^+$ and $\pi^-$ mesons respectively). It should be noted
that a lowering of the threshold for resonance excitation occurs already in the absence of the isospin dependent part of the mean-field due to the
momentum dependence of the optical potential (solid curves in~\figref{delsrt}). The final picture is thus a result of the particular momentum dependence of the optical
and symmetry potentials, the isospin dependence of the latter and the average impact energy of nucleons that suffer 2-body collisions
(which determines the energy at which these potentials are evaluated at). A similar situation is met for the LEC scenario, however in this case
the threshold shift due to the optical potential is half in magnitude (on average 15 MeV vs. 30 MeV) and the dependence of the threshold shift
on asy-Eos stiffness for the $\pi^-$ meson is reversed as compared to the GEC scenario.


\begin{figure}[t]
\centering\includegraphics[width=14pc]{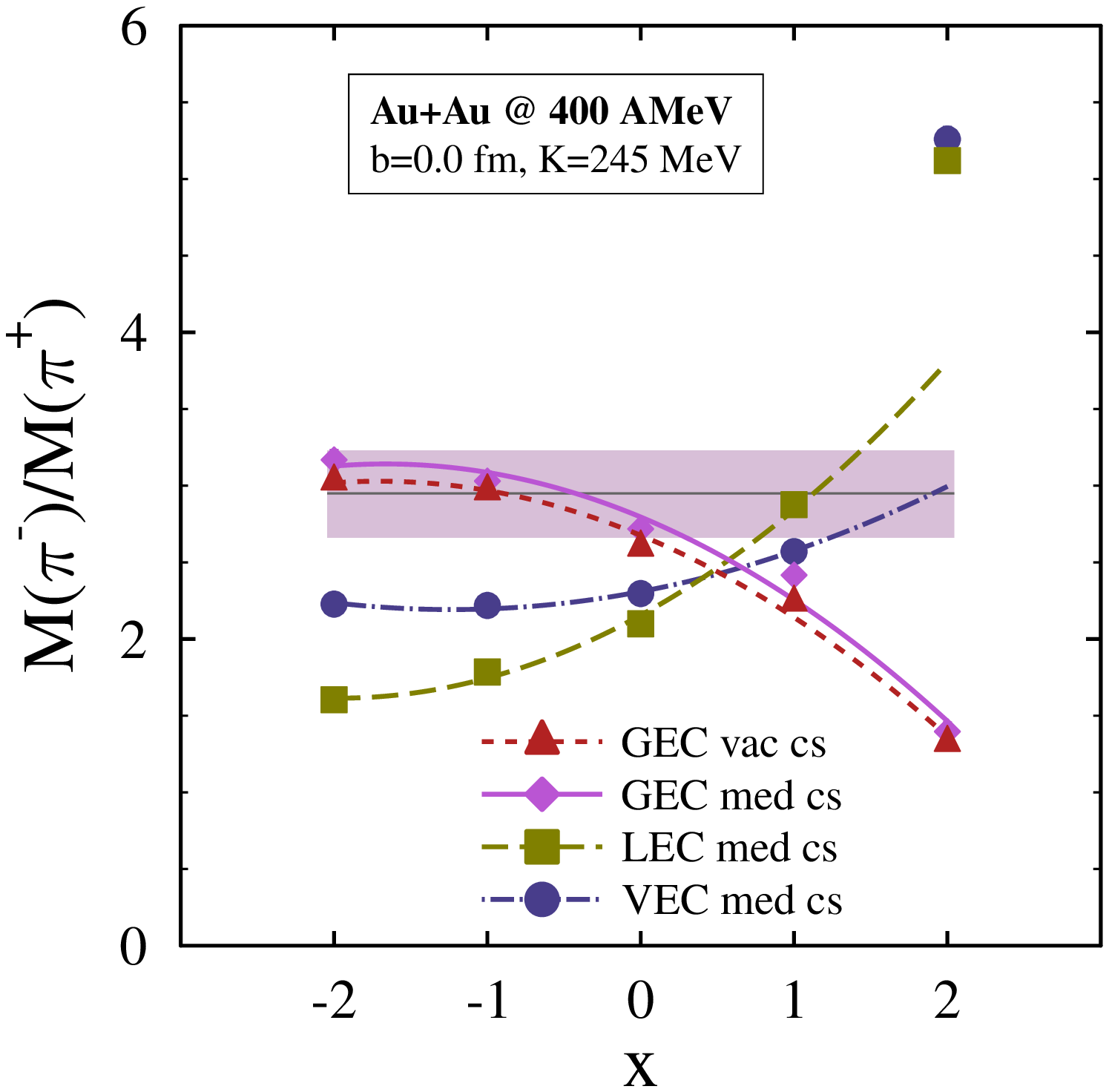}
\caption{\figlab{pionrat_divecs} The same as~\figref{pionmult_divecs} but for the PMR.}
\end{figure}

The PMR theoretical values for the three scenarios described above are compared with the experimental FOPI data, for different
values of the stiffness parameter $x$, in~\figref{pionrat_divecs}. As a consequence of the differences observed for PM, in particular
those of the $\pi^-$ meson, the results for PMR for the GEC (solid curve) scenario differ substantially from those obtained if the either
LEC (dash-dotted curve) or VEC (dashed curve) is enforced. VEC favors an extremely soft asy-EoS and LEC leads to a somewhat 
stiffer one but still in the soft region. Constraints
for the asy-EoS extracted using the GEC scenario range, in view of the magnitude of the experimental uncertainty and the smaller sensitivity
of the PMR to the asy-EoS stiffness in this region, from a stiff to a linear density dependence of SE, compatible at the softer limit
with constraints extracted from nuclear structure measurements. It is noteworthy that constraints extracted from PMR using the GEC
scenario are in complete agreement with those previously extracted from elliptic flow data~\cite{Cozma:2013sja}. The kinetic energy
spectra of PMR show a similar dependence on asy-EoS stiffness and energy conservation scenario over the entire range of accessible
final state pion energies. The power-law parametrization of the symmetry potential has been implemented too in order to perform a comparison 
with the results of Song $\it et~al.$~\cite{Song:2014xza}. Similar results (not shown) as compared to the Gogny parametrization 
are obtained for PMR for both GEC and LEC scenarios, a slightly softer asy-EoS is however favored.


\begin{figure}[t]
\centering\includegraphics[width=14pc]{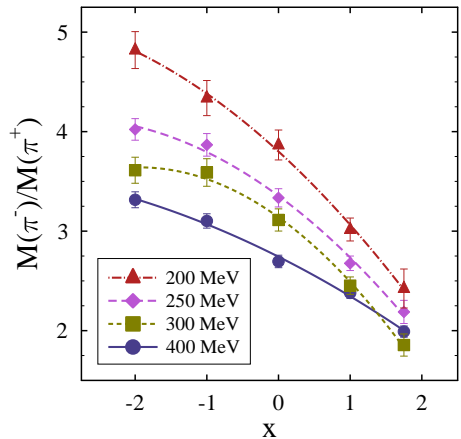}
\caption{\figlab{pionrat_divener} Dependence of the PMR on the asy-EoS stiffness for several values of the impact energy per nucleon in
central $^{197}$Au+$^{197}$Au collisions.}
\end{figure}

The energy dependence of the PMR ratio is addressed in~\figref{pionrat_divener}, where results for projectile impact energies
equal to 200 (dash-dotted curve), 250 (dashed curve), 300 (short-dashed curve) and 400 MeV/nucleon (solid curve) are displayed 
for central (b=0.0 fm) $^{197}$Au+$^{197}$Au collisions. To produce the results
in this figure heavy-ion collisions stopping time has been increased to 100 fm/c, as compared to 60 fm/c for the previously presented
ones, to account for a longer life-time of the high density fireball at lower impact energies. The sensitivity of PMR
to asy-EoS stiffness increases towards lower energies, particularly for impact energies below the vacuum pion production threshold, the
ratio PMR(x=-2)/PMR(x=2) increasing by a factor of 2 between the extreme cases T$_{lab}$=400 MeV/nucleon
and T$_{lab}$=200 MeV/nucleon. Experimental data for impact energies as low a possible are thus clearly desirable.

The results presented up to this point have been obtained with the choices for the isoscalar and isovector baryon potentials
mentioned in Section~\ref{themodel}. In view of the mentioned uncertainties we have studied the impact of variations
of these scenarios on PMR, the results being presented in~\figref{pionrat_divdeltapot}. The $\Delta$(1232) isoscalar
potential has been varied by 25$\%$ around its selected value. The impact on the PMR is moderate, as depicted by the
dark band in~\figref{pionrat_divdeltapot}, a more attractive isoscalar potential resulting in a lower value for PMR and vice versa. 
Additionally, the strength of the isovector $\Delta$(1232) potential has been altered. The $\Delta$ isobar potentials of~\eqref{choicedeltapot}
can be rewritten in the form
\begin{eqnarray}
 \begin{array}{lcr}
V_{\Delta^-}&=& V_N+(3/2)\,V_v\\
V_{\Delta^0}&=&V_N+(1/2)\,V_v\\
V_{\Delta^+}&=&V_N-(1/2)\,V_v\\
V_{\Delta^{++}}&=&V_N-(3/2)\,V_v
\eqlab{choicedeltapot2}
\end{array}
\end{eqnarray}
where $V_N$ is the isoscalar nucleon potential and $V_v$=$\delta$, with the definition $\delta$=(1/3)($V_n$-$V_p$). 
By varying the magnitude of $V_v$ different scenarios for the strength of the isovector baryon potential
can be explored. The results for the choices $V_v$=0 (long-dashed curve), $\delta$ (solid curve), 2$\delta$ (dash-dotted curve)
 and 3$\delta$ (short-dashed curve) have been plotted in~\figref{pionrat_divdeltapot}.
The fourth choice leads, in the case of a momentum independent potential, to no threshold effects. It is thus not surprising that in this case a 
result similar to the VEC scenario is obtained. For the intermediate case $V_v$=2$\delta$ the PMR shows no dependence on the stiffness 
of the asy-EoS, while for the first case, $V_v$=0, which neglects any isospin dependence of the baryon potentials (with the exception
of nucleons), the largest dependence on the asy-EoS stiffness is observed. We conclude that the constraint extracted from PMR for the
asy-EoS stiffness is highly sensitive to the strength of the iso-vector $\Delta$ potential. A trustworthy result cannot thus be obtained 
without a proper knowledge of this quantity and to a lesser extent of the isoscalar $\Delta$ potential.

 It is tempting to extract constraints for the asy-EoS stiffness, for the sake of comparison with results available in the literature, for the standard
scenario of the $\Delta$(1232) potential. Using the results from~\figref{pionrat_divdeltapot}(solid curve) and~\tabref{xvslsymksym} one
arrives at $L$=91$\pm$45 MeV and $K_{sym}$=28$\pm$291 MeV at 1$\sigma$ uncertainty level. The result for the slope parameter $L$ is
in full agreement with constraints extracted from elliptic flow experimental data using any of the TuQMD, UrQMD or IBUU transport models. 
It is noteworthy that for the no isovector $\Delta$ potential case ($V_v$=0, long-dashed curve in ~\figref{pionrat_divdeltapot}) a result which falls on top of nuclear structure measurements is obtained for the
slope parameter:  $L$=61$\pm$22 MeV and additionally $K_{sym}$=-163$\pm$146 MeV. In contrast, a comparison between experimental 
data and the theoretical results for the scenarios with stronger isovector $\Delta$ potentials ($V_v$=2$\delta$, 3$\delta$) suggests
 that they may be unrealistic.

\begin{figure}[t]
\centering\includegraphics[width=14pc]{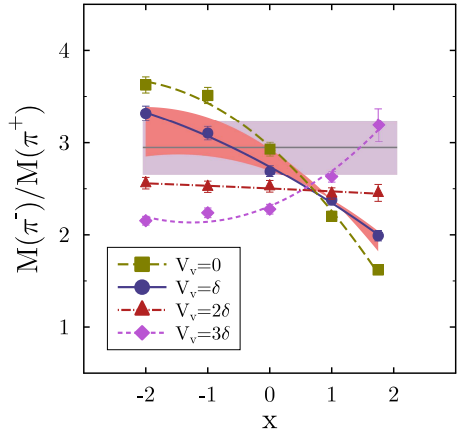}
\caption{\figlab{pionrat_divdeltapot} The impact of the strength of the isoscalar (band) and isovector (curves) in-medium $\Delta$ potentials on PMR
for central $^{197}$Au+$^{197}$Au collisions at a projectile incident energy of 400 MeV/nucleon. The iso-vector $\Delta$ potential $V_v$ is defined
in \eqref{choicedeltapot2}.
}
\end{figure}

We conclude by displaying, in \figref{npflowrat_divecs}, the impact of the three energy conservation scenarios on the neutron-proton 
elliptic flow ratio in $^{197}$Au+$^{197}$Au collisions at a collision energy of 400 MeV/nucleon and integrated impact
parameter b$\leq$7.5 fm (VEC - dashed-dotted, LEC - dashed and GEC - solid curves).
Due to the fact that for elastic collisions the threshold shifts amount, on average, only a few MeV 
(see bottom panel of~\figref{delsrt}), the impact of LEC and respectively GEC on elliptic flow values of
neutrons and protons is small, resulting in a slight impact on the extracted constraints for asy-EoS from this observable.
An identical statement holds true for neutron-proton elliptic flow differences. There is however a noticeable impact on flows
in the case when medium effects on cross-sections are introduced for the elastic channels below pion production threshold. 
The impact on neutron-proton elliptic flow ratio is also in this case within the uncertainty
of the extracted asy-EoS constraints, however the impact on neutron-proton elliptic flow differences is larger. 
The density and asymmetry dependence of elastic and inelastic nucleon-nucleon cross-section
is thus a topic that necessitates further study. We can however conclude that for the GEC scenario compatible constraints for
the asy-EoS stiffness can be extracted from pion multiplicity and elliptic flow observables.

The presented model lacks a few ingredients that may prove important. First on the list is the in-medium pion potential,
which may have an important impact on pion multiplicities close to threshold~\cite{Hong:2013yva} given its estimated 
S wave magnitude from chiral perturbation theory~\cite{Kaiser:2001bx} or from the explanation of the existence of pionic atoms~\cite{Itahashi:1999qb}. 
Within the scenario of global energy conservation two-body collisions lose their local character and effects such as
retardation may be of relevance. Such an expectation is supported by the fact that the ratio  of the size of the high density
nuclear  matter fireball and its lifetime is of the order of the speed of light. Thirdly, given the size of the threshold shifts of
pion production threshold of a few tens of MeV and the sensitivity of the PMR to small variations, it may be also expected that, for accurate
predictions, isospin breaking effects in the $\Delta$(1232) isospin quadruplet (both mass and decay width) could have an impact, even though a
secondary one. Work to include these presently omitted contributions is planned.

It would be of interest to extend the current model to include the strangeness production channels and reexamine
~\cite{Fuchs:2000kp,Hartnack:2005tr} or extract constraints of the high density dependence of the isoscalar and isovector components of the
nuclear matter equation of state using presently or in the near future available experimental information 
collected by the KaoS~\cite{Sturm:2000dm}, FOPI~\cite{Lopez:2007rh} and HADES~\cite{Pietraszko:2014pca} Collaborations from GSI.

\begin{figure}[t]
\centering\includegraphics[width=14pc]{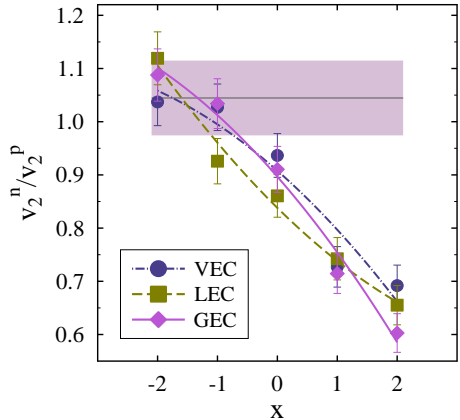}
\caption{\figlab{npflowrat_divecs} The impact of the different energy conservation scenarios on the neutron-proton
effective flow ratio. The horizontal band represents the experimental FOPI-LAND results listed in Ref.~\cite{Cozma:2013sja}. Using
one centrality bin for the experimental data analysis results in a lower value, by about 3$\%$, for the neutron-proton elliptic flow ratio
~\cite{Tra:2014aaa}. Preliminary results of the ASYEOS Collaboration favor a lower value for this observable too~\cite{Russotto:2014zba}.}
\end{figure}

\section{Conclusions}
An upgraded version of the T\"{u}bingen QMD transport model, which allows the conservation of energy in intermediate energy
heavy-ion collisions at an event by event basis, has been presented. The conservation of energy in two-body collisions has been implemented at
local and, alternatively, at global level by taking into account in the energy conservation constraint for two-body collisions of the in-medium 
potential energies of the propagating particles, which has as a result
the shift of particle production thresholds and effective energies at which elastic collisions take place inside dense nuclear matter.
It has been shown that the impact of such effects on pion multiplicities and ratios is important when energy conservation
is enforced at a global level and only moderately so for the local scenario. This results in very different extracted constraints for the
asy-EoS stiffness using the vacuum, local and global energy conservation scenarios. 

The impact on elliptic flow values of neutrons and
protons is however small due to threshold shifts that amount on average only a few MeV, an order of magnitude smaller
than in the pion production case. In the case of the global energy conservation scenario an almost perfect agreement between
the constraints on the asy-EoS stiffness extracted from pion multiplicity and elliptic flow observable is achieved, compatible
with a linear dependence on density of the symmetry energy above the saturation point.

 An increasing sensitivity of the pion multiplicity ratio to the stiffness of the asy-EoS towards lower incident energy collisions is observed, in agreement with
expectations. The effect increases by a factor of two at T$_{lab}$=200 MeV/nucleon with respect to the lowest energy point for which experimental
data for pion multiplicities are available, T$_{lab}$=400 MeV/nucleon. Measurements at incident energies around and below pion production threshold
are thus extremely desirable and promising.

The impact of the poorly known $\Delta$(1232) in-medium potentials on the pion multiplicity ratio has been investigated.
The impact of the isoscalar part when its strength is modified by 25$\%$ from the usual one, equal to that of the nucleon,
is moderate. The pion multiplicity ratio is however demonstrated to be highly sensitive to the isovector part of the $\Delta$(1232)
potential, constraints for the asy-EoS stiffness ranging from very stiff to extremely soft can be extracted depending on the choice made
for this quantity. The observed sensitivity hinders at present the extraction of meaningful constraints for the density dependence
of the symmetry energy above saturation using the charged pion multiplicity ratio, in spite of its proven dependence on it.

 Progress on the theoretical side, by extracting information on the isovector part of the $\Delta$(1232) potential either from ab-initio calculations or from
phenomenological descriptions using the same transport model for reactions in which it plays a role (e.g. pion-nucleus scattering) can
alleviate the noted problem. In the worst case scenario the realistic isovector $\Delta$ potential may turn out to be such
that it cancels out the sensitivity of the pion multiplicity ratio to the stiffness of the asy-EoS. Even in such a case, the experimental
measurement of this observable will be extremely valuable since, due to the many effects that impact it,
it will allow a precise test of our understanding of the hadronic interactions in the intermediate energy range.

Further work to include so far neglected potentially important contributions, 
like the in-medium pion potential, retardation effects and possibly even isospin breaking effects in the $\Delta$(1232)
isospin quadruplet, and an extension of the model to include strangeness production channels is planned.

\section{Acknowledgments}
The research of M.D.C. has been financially supported by the Romanian Ministry of Education and
Research through contract PN09370102. The author would like to thank Prof. Dr. Wolfgang Trautmann for a careful reading of the manuscript and useful
suggestions. The assistance of the DFH and DFCTI departments of IFIN-HH with the maintenance of the computing cluster on which simulations
were performed is gratefully acknowledged.

\bibliographystyle{model1a-num-names}
\bibliography{references}

\end{document}